\documentclass[letterpaper,superscriptaddress,aps,twocolumn]{revtex4-2}
\usepackage{graphicx,color}
\usepackage{todonotes}
\usepackage{epstopdf}
\usepackage{amssymb}
\usepackage{multirow}
\usepackage{subcaption}
\usepackage{dcolumn}
\usepackage{bm}
\usepackage{amssymb,amsmath} 

\usepackage{siunitx}
\definecolor{strawberry}{rgb}{1.0,0.0,0.5}

\begin{document}
\preprint{APS/123-QED}

\title{Relaxation mechanisms in supercooled liquids past the Mode--Coupling Crossover: Cooperatively Re--arranging Regions vs Excitations}

\author{Levke Ortlieb}
\affiliation{H.H. Wills Physics Laboratory, Tyndall Avenue, Bristol, BS8 1TL, UK}

\author{Trond S. Ingebrigtsen}
\affiliation{DNRF Centre for Glass and Time, IMFUFA, Department of Science, Systems and Models, Roskilde University, Postbox 260, DK--4000 Roskilde, Denmark}

\author{James E. Hallett}
\affiliation{H.H. Wills Physics Laboratory, Tyndall Avenue, Bristol, BS8 1TL, UK}

\author{Francesco Turci}
\affiliation{H.H. Wills Physics Laboratory, Tyndall Avenue, Bristol, BS8 1TL, UK}

\author{C. Patrick Royall}
\affiliation{H.H. Wills Physics Laboratory, Tyndall Avenue, Bristol, BS8 1TL, UK}
\affiliation{Gulliver UMR CNRS 7083, ESPCI Paris, Universit\'{e} PSL, 75005 Paris, France}
\affiliation{School of Chemistry, University of Bristol, Cantock’s Close, Bristol, BS8 1TS, UK}

\date{\today}

\begin{abstract}
Among the challenges in discriminating between theoretical approaches to the glass transition is obtaining suitable data. In particular, particle--resolved data in liquids supercooled past the mode--coupling crossover has until recently been hard to obtain. Here we combine nano-particle resolved experiments and GPU simulation data which addresses this and investigate the predictions of differing theoretical approaches. We find support for both dynamic facilitation and thermodynamic approaches. In particular, \emph{excitations}, the elementary units of relaxation in dynamic facilitation theory follow the predicted scaling behaviour 
and the properties of \emph{cooperatively rearranging regions} (CRRs) are consistent with RFOT theory. At weak supercooling there is limited correlation between particles identified in excitations and CRRs, but this increases very substantially at deep supercooling.  We identify a timescale related to the CRRs which is coupled to the structural relaxation time and thus decoupled from the excitation timescale, which remains microscopic.
\end{abstract}

\maketitle

\textit{Introduction --- .}
Understanding the origin of the glass transition is a longstanding challenge in condensed matter physics. Cool or compress any liquid fast enough and it will not order into a crystal but remain a liquid before, eventually, it falls out of equilibrium and becomes a \emph{glass}. Over the years, a variety of theoretical approaches have been developed to account for the massive slowdown in the dynamics of such supercooled liquids as they approach the experimental glass transition temperature $T_g$~\cite{berthier2011,ediger2012,berthier2016}. What happens below $T_g$ remains a matter of conjecture, as equilibration is not possible, with wildly differing theoretical standpoints which nevertheless provide equally good interpretations of the available data~\cite{berthier2011,hecksher2008}. These can be broadly classified into those which presume that the dynamic slowdown in supercooled liquids is predominantly a \emph{dynamic} phenomenon, and those which invoke a \emph{thermodynamic} origin~\cite{berthier2011,ediger2012,berthier2016}.

Among the former is \emph{dynamic facilitation} \cite{speck2019,chandler2010}, in which the arrest is attributed to the emerging kinetic constraints in the liquid. This approach focuses on \emph{dynamic heterogeneities}~\cite{berthier}, i.e., the coexistence of mobile and solid-like regions in a supercooled liquid. Dynamic facilitation is built around the idea of a \emph{dynamical} phase transition. Here, unlike conventional thermodynamic phase transitions, where the coexisting phases are characterised by distinct static properties (e.g., the density difference in liquid-gas coexistence), the dynamical transition amounts to a coexistence of mobile regions where the particles relax on a microscopic timescale and immobile regions where relaxation is very slow. The nature of the mobile regions is characterised by localized \emph{excitations} which constitute the elementary units of relaxation in dynamic facilitation. Their dynamics are governed by kinetic constraints, but have recently been related to local structure ~\cite{hasyim2021}.

\begin{figure*}
{\includegraphics[width=0.99\textwidth]{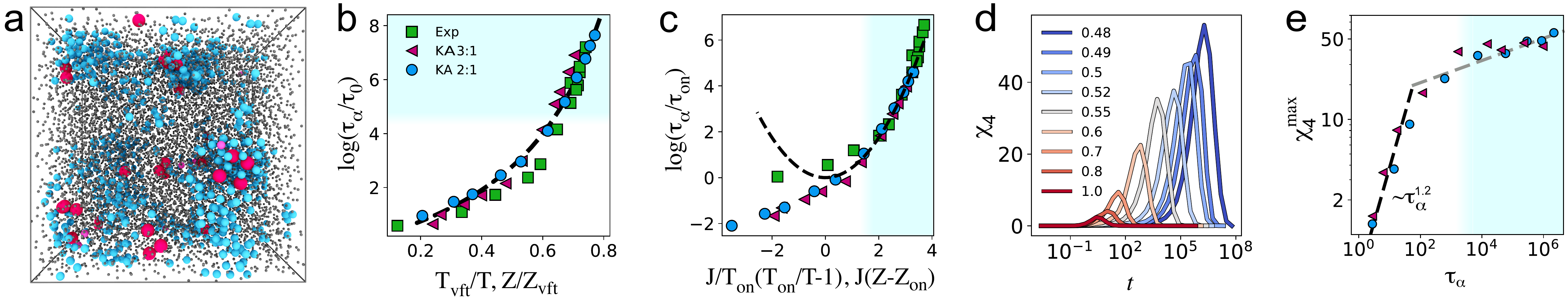}}
\caption{
Relaxation well below the mode--coupling crossover.
(a) Snapshot of the 2:1 KA model at $T=0.48$. Bright pink: particles in both CRRs and excitations. Blue: CRR. Pale pink: excitations. Grey: neither. Particles are not rendered to scale.
(b) Angell plot with scaled VFT parameters (Eq. (\ref{eqVFT})). The dashed line is the VFT fit for KA 2:1. The legend in (b) pertains also to (c) and (e).
(c) Angell plot scaled with parameters of the parabolic fit (Eq. (\ref{eqParabolic})). The dashed line is the parabolic fit for KA 2:1.  
(d) Dynamic susceptibility $\chi_4(t)$ for different $T$ for the KA 2:1 model. 
(e) maximum of $\chi_4$ vs. $\tau_{\alpha}$, compared with the $\tau_{\alpha}^{1.2}$ scaling seen previously \cite{coslovich2018,berthier2017}  (black dashed line). Grey dashed line is a guide to the eye.
Blue shading indicates temperatures or relaxation times past the mode coupling temperature where conventional simulations are limited. 
}
\label{figAngell}
\end{figure*}

By contrast, a number of theoretical approaches have been developed which emphasise the role of \emph{thermodynamical properties}, expressed through the structure assumed by the constituent particles. While the two-point structure, (i.e., the pair correlation function) changes little, higher-order correlations exhibit a marked change upon approaching the glass transition \cite{royall2015}. In particular, Adam-Gibbs theory and Random First-Order Transition (RFOT) theory emphasise configurational entropy and the role of so-called co-operatively re-arranging regions (CRRs). These constitute groups of particles with access to only a few distinct states. For a fixed number of states per CRR, the drop in configurational entropy means that CRRs are expected to grow in size as the system approaches the glass transition. In fact, the entropy with respect to an ideal gas has also been shown to play an important role in understanding supercooled liquids \cite{bell2020}. Other theories, such as geometric frustration, emphasize \emph{locally favoured} -- or locally \emph{preferred -- structures} (LFS) \cite{tarjus2005}. These locally favoured structures are geometric motifs that are minima of the local (free) energy. Their concentration increases as a glass-former is cooled, and they have been identified with the emergence of slow dynamics~\cite{royall2015physrep,coslovich2007,royall2008,leocmach2012} and a drop in configurational entropy \cite{hallett2018,hallett2020}.

Among the challenges of identifying which theoretical approach best describes the glass transition is obtaining equilibrated data in a suitable temperature range (or, in the case of hard sphere--like systems, a suitable density range). In particular, one may identify four important temperatures (or densities): Firstly, $T_\mathrm{on}$ ($\tau_\alpha(T_\mathrm{on})\sim\tau_\mathrm{on}$ ) the onset temperature where upon cooling, the system begins to exhibit slow dynamics. Secondly, the temperature of the mode-coupling crossover, $T_\mathrm{mct}$ ($\tau_\alpha(T_\mathrm{mct})\sim10^4\tau_0$). Here $\tau_0$ is a microscopic timescale. Thirdly the operational glass transition temperature where the structural relaxation time of molecular glassformers exceeds 100s $T_\mathrm{g}$ ($\tau_\alpha (T_\mathrm{g})\sim10^{14}\tau_0$). Finally, the Kauzmann temperature $T_K$ where the entropy of a 
supercooled liquid would be extrapolated to drop below the crystal entropy, which may correspond to a thermodynamic transition to a state with vanishing configurational entropy and divergent structural relaxation time.

For some time, computational and experimental techniques which could provide particle--resolved data capable of delivering the kind of measurements (many-body spatial and temporal correlation functions) that might discriminate between the theoretical interpretations were limited to $T\gtrsim T_\mathrm{mct}$ for computer simulations and volume fraction $\phi\lesssim\phi_\mathrm{mct}$ for particle-resolved studies of colloidal hard-sphere like systems \cite{hunter2012,ivlev}. This is a significant issue as theories such as RFOT predict a quantitative change in the mechanism of relaxation at the mode-coupling crossover, which may be interpreted as the high-temperature limit of stability of the glassy state \cite{biroli2014}.

Recently, this has begun to change, with the advent of SWAP Monte Carlo computer simulations, which generates particle-resolved equilibrated configurations at temperatures even below $T_\mathrm{g}$ \cite{berthier2017}. However, this technique is limited when it comes to distinguishing the \emph{dynamic} predictions of certain theories. While theories which predict static quantities (such as RFOT) may be tested (with some success) \cite{berthier2017,ninarello2017,gokhale2016}, those theories based on dynamical approaches such as dynamical facilitation have only been investigated in the regime between the onset temperature and the mode-coupling crossover $T_\mathrm{on} \gtrsim T \gtrsim T_\mathrm{mct}$ \cite{keys2011}. Again, some success of the predictive power of dynamic facilitation was found \cite{keys2011,gokhale2014}, but it is notable that the mode-coupling theory also provides an accurate description in this temperature range ~\cite{janssen2018}. Obtaining particle--resolved data for more deeply supercooled systems is thus necessary to address the challenge of identifying the 
most successful approach in describing the glass transition, which forms the focus of the present work.

\begin{figure*}
\centering
{\includegraphics[width=0.98\textwidth]{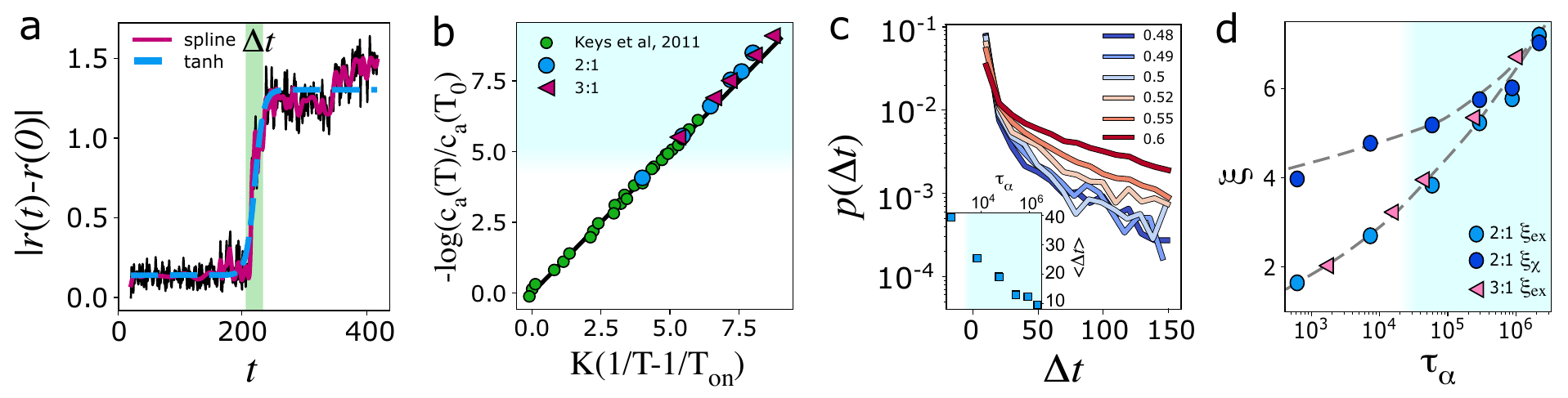}}
\caption{Testing the predictions of dynamic facilitation.
(a) Single-particle displacement to illustrate the algorithm to identify excitations (see SM for details). 
(b) Scaled temperature dependence of the concentration of excitations in comparison with the data by Keys et al. \cite{keys2011}. 
(c) Probability distribution of excitation durations $\Delta t$. Inset: mean of the $P(\Delta t)$ distribution.
(d) Excitation lengthscale $\xi_\mathrm{ex} = (c_a/t_\mathrm{traj})^{-1/d}$ and lengthscale inferred from the maximum of $\chi_4$, $\xi_{\chi}$. 
Grey dashed lines are guides to the eye.
\label{figFacilitation}}
\end{figure*}

Here we exploit two new techniques to obtain dynamic data in the regime beyond the mode-coupling crossover $T<T_\mathrm{mct}$ to probe the predictions of thermodynamic and dynamic theories. The first of these developments is nano-particle resolved studies of colloidal systems \cite{hallett2020}, which by using smaller colloids takes advantage of the timescale of colloidal systems $\tau_B\sim\sigma^3$, where  $\sigma$ is the particle diameter. By using smaller particles to effectively accelerate the system, for a given measurement time $>\tau_\alpha$, the ratio $\tau_\alpha/\tau_B$ can be increased such that data can be obtained in the $\phi\gtrsim\phi_\mathrm{mct}$ regime.  The second new development is highly-efficient GPU simulations using the Roskilde University Molecular Dynamics (RUMD) package  \cite{bailey2017}. Both techniques deliver dynamic data for state points where the structural relaxation time is around a thousand times that at the mode-coupling crossover \cite{hallett2020,ingebrigtsen2019}.

We use these developments to investigate which theoretical approach -- dynamic or thermodynamic -- best describes our data. In particular, we enquire, can we identify CRRs and do they become more compact as the temperature is reduced, as predicted by RFOT? Additionally, can we identify excitations? Do they obey the assumptions of dynamic facilitation, in that their nature remains essentially unchanged, but their population drops in an Arrhenius-like manner upon supercooling? Identification of CRRs and excitations then allows us to enquire -- are these relaxation mechsnisms predicted by such profoundly different theoretical standpoints, in fact different facets of the same phenomena? That is to say, to what extent do the same particles constitute CRRs and excitations?

To proceed, we perform nano-particle resolved experiments \cite{hallett2018,hallett2020} of hard-sphere-like colloids and computer simulations of the binary Lennard-Jones Kob-Andersen (KA) mixture. The simulations are performed at compositions of 2:1 and 3:1, as the often-used 4:1 composition has proven unstable against crystallisation, while the 2:1 and 3:1 compositions are 
stable to crystallisation on the timescales we consider \cite{ingebrigtsen2019}. A snapshot of the 2:1 KA system at temperature $T=0.48$, the lowest temperature under study, with $N = 10,002$ is shown in Fig. \ref{figAngell}(a). Unless otherwise specified, we present data from this composition and system size. Experiment and computer simulation details are in the SM.

\emph{Dynamical particle--resolved data at deep supercooling --- .}
In Fig. \ref{figAngell}(b), we present a so-called Angell plot of the structural relaxation time $\tau_\alpha$ (determined with a stretched exponential fit to the self 
intermediate scattering function, see SM). To compare the simulations and experiments in the same plot, we use $1/T$ and the compressibility factor $Z$, respectively, and scaled with the temperature from fit to the Vogel--Fulcher--Tamman (VFT) equation:
\begin{equation}
\tau_{\alpha} = \tau_{0}\exp\left(\frac{DT_{\rm vft}}{T-T_{\rm vft}}\right)
\label{eqVFT}
\end{equation}

\noindent
where $\tau_0$ is a microscopic timescale and $D$ represents the fragility.
In Fig. \ref{figAngell}(b), the compressibility factor $Z$ is determined from the volume fraction $\phi$ via the Carnahan--Starling relation \cite{hallett2018,berthier2009}.

We also show the collapse of the dynamical data in the parabolic law of dynamic facilitation in Fig. \ref{figAngell}(c). 
\begin{equation}
\tau_{\alpha} = \tau_0\cdot\exp\left\{\frac{J^2}{T_{\rm on}^{2}}\left(\frac{T_{\rm on}}{T}-1\right)^2\right\}
\label{eqParabolic}
\end{equation}

These plots reveal the new dynamical regime accessible to our techniques. As has been seen previously \cite{elmatad2009,hecksher2008}, both facilitation and the VFT equation (which has been related to thermodynamic interpretations \cite{cavagna2009}) fit the data essentially equally well. In the SM we show fits to the predictions of MCT (Fig. \ref{sFigMCT}) and detail the parameters resulting from the various fits in table \ref{sTableParameters}. 

We also show the dynamic susceptibility $\chi_4(t)$ in Fig. \ref{figAngell}(d), whose peak value is a measure of the number of particles involved in dynamically heterogeneous regions. When we plot the peak,  $\chi_4^\mathrm{max}$ as a function of the relaxation time [Fig. \ref{figAngell}(e)], we see behaviour consistent with a crossover around the mode--coupling temperature $T_\mathrm{mct} \approx 0.55$  and a slower growth at deeper supercooling. This is expected in the RFOT context of a change in relaxation behaviour around $T_\mathrm{mct}$. 
For weaker supercooling $T>T_\mathrm{mct}$, we see behaviour consistent with the scaling $\chi_4^\mathrm{max} \sim \tau_\alpha^{1.2}$ as obtained previously \cite{coslovich2018}.


\emph{Identifying excitations --- .}
We now turn to analysing these data in such a way that we might discriminate between the dynamical and thermodynamic approaches. To do so, first we describe a procedure to identify particles in excitations, which is related to that described in \cite{keys2011}.

We set a timescale $t_a = 200$ and a length scale $a=0.5$  (both in Lennard-Jones units) and a trajectory of length $t_\mathrm{traj} = 5 t_a$ is generated from equilibrated configurations. To ensure that the particle commits to the new position, the average position of an eligible particle in the first and the last $t_a$ of the trajectory is required to be $>a$. To identify when the excitation happened, we check in each frame if the average position of the previous and the following $t_a$ are at least $a$ apart. The time at which the excitation occurs and its duration $\Delta t$ are chosen from a hyperbolic tangent fit of the displacement after smoothing the trajectory with a spline [see Fig. \ref{figFacilitation}(a)] (see SM for further details). Clearly such a set of criteria selects a \emph{subset} of mobile particles.

We express the population of excitations 
as the percentage of particles identified in excitations during a trajectory of length $t_\mathrm{traj}$, $c_a$. We find the expected scaling of the population of particles in excitations, that is to say $c_{a} \sim \exp[K(1/T-1/T_\mathrm{on})]$ in Fig. \ref{figFacilitation}(b) which is consistent with Ref. \cite{keys2011}. Here, the energy scale related to the excitation population $K_{2:1}=9.6$ and  $K_{3:1}=15.5$.

An important property of excitations predicted by facilitation is that their duration and spatial extent remains unchanged during cooling. In Fig. \ref{figFacilitation}(c) inset, we see that the mean duration of excitations drops upon supercooling, although the magnitude of the drop is less than one order of magnitude. Turning to the distribution of durations of excitations in Fig. \ref{figFacilitation}(c), for temperatures well below $T_\mathrm{mct}$($\approx$ 0.55 for KA 2:1) the distributions lie almost on top of one another. At higher temperature, $T \gtrsim T_\mathrm{mct}$, we see a longer tail in the distribution of $P(\Delta t)$. This change may be perhaps related to the different nature of relaxation at higher temperature, approaching the mode-coupling crossover and at temperatures above it
\cite{biroli2014}.

\begin{figure}
\includegraphics[width=\linewidth]{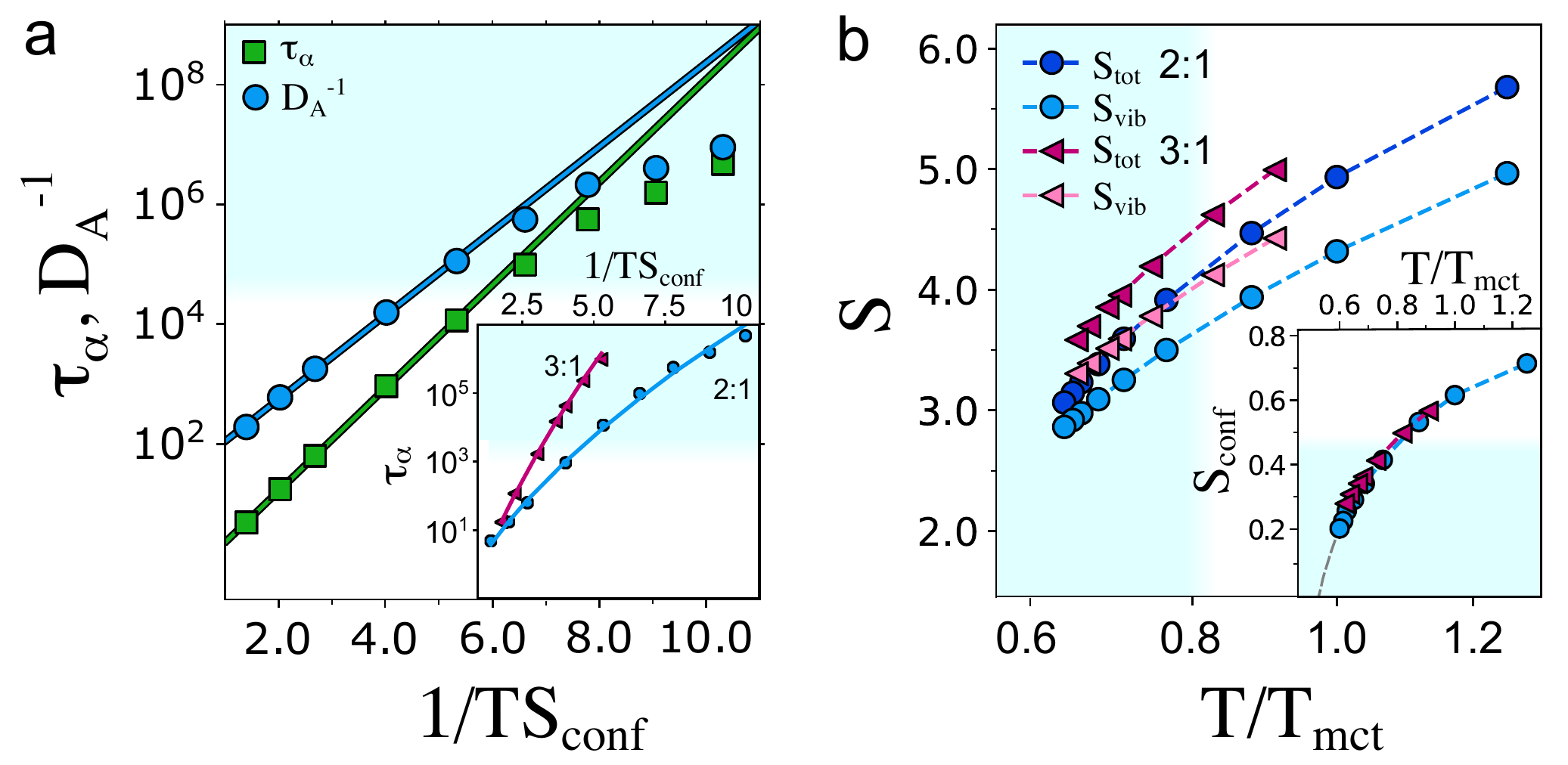}
\caption{Entropy scaling in the Adam-Gibbs model and RFOT. 
(a) The relaxation time $\tau_{\alpha}$ and diffusion coefficient $D_{\rm A}^{-1}$ as functions of $1/TS_{\rm conf}$. The Adam-Gibbs relation
breaks down below $T_{\rm mct}$.
Inset: Plot of the RFOT prediction Eq. (\ref{eqRFOTFit}) with the exponent $\alpha$ as a best fit.
(b) The total entropy $S$ and vibrational entropy $S_{\rm vib}$  as a function of $T/T_{\rm mct}$.
The inset shows the configurational entropy $S_{\rm conf}$. Grey dashed line is an extrapolation. 
\label{figAdamGibbs}}
\end{figure}

Excitations are also expected to remain small which contrasts with CRRs which are expected to grow. Particles in excitations at the lowest temperature sampled $(T=0.48)$ are shown in pink in Fig. \ref{figAngell}(a) and are identified as single isolated particles. In Fig. \ref{sFigSnapshotT055} we show contrasting data at T=0.55 $(\approx T_\mathrm{mct})$. At the higher temperature, there are many more particles in excitations. Thus at the level of our analysis we find that the the vast majority at least 92\% and approaching 100\% at low temperature of excitations are single particles.

We can also define a dynamic lengthscale from the population of excitations as $\xi_\mathrm{ex} = (c_a/t_\mathrm{traj})^{-1/d}$. In Fig. \ref{figFacilitation}(d) we compare this with a lengthscale of dynamic heterogeneity inferred from the dynamic susceptibility, $\xi_\mathrm{\chi} = (\chi_4^\mathrm{max})^{1/d}$. These lengths show distinct scaling for most of the temperature regime that we access, but it is possible that, at lower temperatures 
they may scale together.

\begin{figure*}
\includegraphics[width=0.99\textwidth]{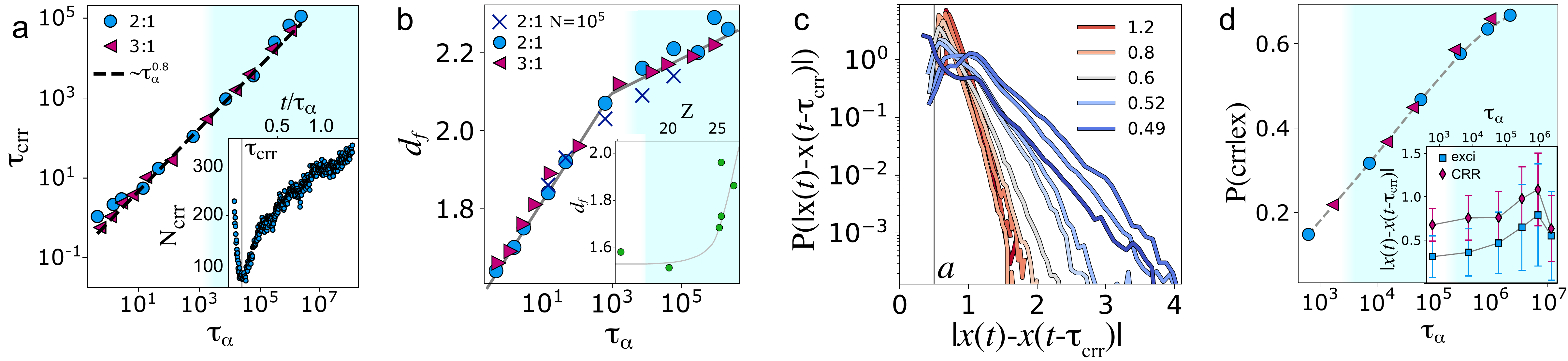}
\caption{Characteristics of CRRs at supercooling past the mode--coupling crossover.
(a) The CRR timescale as a function of the relaxation time. Inset: The timescale on which the number of CRRs is minimised, $\tau_\mathrm{crr}$.
(b) Fractal dimension of CRRs. Inset experiment measurements of the CRR fractal dimension.
(c) Distributions of mobilities of particles in CRRs.
(d) change of the probability that a particle that is in an excitation is also part of a CRR with $\tau_{\alpha}$.
Grey dashed lines are guides to the eye.
\label{figCRR}.}
\end{figure*}

\emph{Thermodynamic interpretation --- .}
We now turn to the configurational entropy approach as emphasized in the Adam-Gibbs relation and RFOT. The activation energy, in the Adam-Gibbs model, is assumed proportional to the volume (size) of CRR regions. The size of these regions can in turn be related to the inverse of the configurational entropy (a measure of the number of distinct local potential-energy minima) defined as $S_{\rm conf} = S - S_{\rm vib}$, where $S_{\rm vib}$ is the vibrational entropy of the liquid. The Adam-Gibbs relaxation time is then given by
\begin{equation}
\tau_{\alpha}  = \tau_{0}\exp(A_{\rm ag}/TS_{\rm conf})
\end{equation}
Figure \ref{figAdamGibbs}(a) shows that the Adam-Gibbs relation holds well on approaching $T_{\rm mct}$ but gradually starts to break down at deeper supercoolings in agreement with other studies \cite{ozawa2019} (see SM for details on the calculations). A breakdown of Adam-Gibbs in two- or four dimensions has been observed before, and in three dimensions has been related to the breakdown of the Stokes-Einstein relation for the viscosity and diffusion coefficient \cite{shila,sengupta2012,parmar2017}. Nevertheless, here we find that for both the relaxation time and the diffusion coefficient that the Adam-Gibbs relation does not hold.

Figure \ref{figAdamGibbs}(b) shows the total entropy, vibrational entropy, and configurational entropy as a function of $T/T_{\rm mct}$.  We find that the configurational entropy 
is extrapolated to vanish at a non-zero temperature $T_{\rm K}$, compatible with previous work \cite{berthier2017}.  We emphasise that such extrapolation is in no sense conclusive, and that other scenarios are possible, such as an avoided transition \cite{stillinger2001,royall2018jpcm}.

RFOT has been shown to work well over a large temperature range \cite{ozawa2019}. We present in Fig. \ref{figAdamGibbs}(a) inset the RFOT prediction that

\begin{equation}
\tau_{\alpha} =  \tau_{0}\exp(A_{\rm rfot}/TS_{\rm conf}^{\alpha}).
\label{eqRFOTFit}
\end{equation}
We treat the exponent $\alpha$ as a fitting parameter and find good agreement with RFOT with $\alpha \approx 
0.61$ and $0.62$ for KA 2:1 and 3:1, respectively. Hence, these 
results do not allow us to distinguish between the two approaches, other than the limitations of Adam-Gibbs theory.

\emph{Cooperatively rearranging regions --- .}
We now seek to identify the CRRs, where we consider the fastest 10\% of particles  to comprise the cooperatively rearranging regions, which we identify via connected regions of mobile particles, with connectivity defined by a bond length of $1.3\sigma$. When we plot the number of CRRs as a function of time (see Fig. \ref{figCRR}(a) inset), we identify the timescale of maximum CRR size, characterised by the minimum number of CRRs. We find that this timescale departs from the structural relaxation time and at deep supercooling, we see that CRRs last for a much shorter time than the relaxation time, and fit the CRR time $\tau_\mathrm{crr} \sim \tau_\alpha^{0.8}$, (see Fig. \ref{figCRR} (a)).

We also determine the fractal dimension of the CRRs (see Fig. \ref{sFigFractal}) which is predicted by RFOT to increase for low temperatures $T\lesssim T_\mathrm{mct}$ as the CRRs become more compact. In Fig. \ref{figCRR}(b) we provide evidence for such a crossover in the fractal dimension of the CRRs. This occurs around the mode--coupling crossover, however it is clear that the increase in fractal dimension must ultimately be limited as $d_f\leq3$. A comparable trend is shown in experimental data, Fig. \ref{figCRR}(b) inset. In trying to obtain a CRR lengthscale, we note that except for percolation the CRR size distribution exhibits a power--law scaling $P(S_\mathrm{crr}) \sim S_\mathrm{crr}^{-2}$ (Fig.  \ref{sFigDistributionNCRR}).

\emph{What does it all mean? --- .}
Finally, we address the question with which we opened this article: which interpretation should we believe? We have shown evidence in support of \emph{both} dynamical and thermodynamic  interpretations, however we can do one further piece of analysis: that is to enquire, are the particles in CRRs and excitations in fact the \emph{same} particles? Such a correspondence would be evidence for these approaches being different ways of describing -- or identifying -- the same relaxing particles. Now the population of particles in excitations becomes rather low at low temperature, but the CRRs have the same population, 10\%, by construction. Moreover, the timescale of the excitations decreases somewhat Fig. \ref{figFacilitation}(c) inset, while that of the CRRs is slaved to the relaxation time, or CRR time. Nevertheless, we enquire of the probability that a particle in an excitation is also in a CRR. As we see in Fig. \ref{figCRR}(d), this increases very markedly at low temperature. So we conclude that, at least to some extent, the excitations of dynamic facilitation and the CRRs of Adam-Gibbs/RFOT could be interpreted as differing ways of identifying the same particles.

Now the excitations are identified in trajectories of $t_\mathrm{traj}=1000$ time units. If we consider the chance that a particle is an excitation on the \emph{much longer} timescale of $\tau_\alpha$, then we estimate  $c_a \tau_\alpha /  t_\mathrm{traj}$ excitations during the relaxation time. For $T=0.55\approx T_\mathrm{mct}$, we estimate 3716 particles to have been in excitations on the duration of the relaxation time, which is much more than 10\% of the system corresponding to particles in CRRs. At the lowest temperature we consider, this grows to 58592 particles, much more than the whole system! We conclude that particles in CRRs have presumably been in  multiple excitations during the relaxation time. In Fig. \ref{figCRR}(c), we show the distribution of how much the CRR particles have moved. Clearly most have moved more that the lengthscale of the excitations $a=0.5$. Now the choice of $a=0.5$ is arbitrary (within certain constraints). In the SM we consider setting $a=1.0$ (see Figs. \ref{sFigExRaw} and \ref{sFigExa1}). Of course, fewer particles are found in excitations, but the scaling is the same and at low temperatures, the number of particles found to participate in an excitation during $\tau_\alpha$ at $T=0.48$ still massively exceeds the 10\% corresponding to particles in CRRs. We conclude that the choice of $a=0.5$ does not qualitatively change our conclusions. We also tried changing the criterion for particles in CRRs to 8\% and 12\% and also found qualitatively similar behaviour.

We therefore conclude that our data provide support for both dynamic facilitation and theories based on thermodynamics. Dynamical facilitation in systems with non--trivial thermodynamics has been related to a dynamical phase transition with a finite--temperature lower critical point \cite{turci2017prx,royall2020}. This is distinct from the zero--temperature termination of the dynamical phase transition in kinetically constrained models without meaningful thermodynamics \cite{chandler2010}.

By probing the predictions of dynamic facilitation and thermodynamic interpretations of the glass transition in previously inaccessible dynamical regimes, we have identified the following:
\begin{itemize}
\item{There is a crossover in the rate of increase of the dynamic susceptibility, which is consistent with a changeover of relaxation mechanisms around $T_\mathrm{mct}$.}
\item{The predictions of dynamic facilitation, in terms of the population of excitations and their nature remained unchanged at low temperature are largely upheld. The mean duration of excitations $\langle \Delta t \rangle$ decreases by a factor of around four over the temperature range we consider. However distribution of excitation timescales appears to show little dependence on temperature for $T<T_\mathrm{mct}$. We identify excitations as single particles (Fig. \ref{sFigSnapshotT055}) and in this sense the spatial extent of excitations is temperature independent.}
\item{The predictions of Adam--Gibbs theory are not upheld. We find significant breakdown of Stokes--Einstein in our systems and hence Adam-Gibbs cannot be correct for both relaxation time and diffusion, but we find no evidence for making this distinction as in other studies. Our data is consistent with the RFOT scaling for $\tau_\alpha$ and $S_\mathrm{conf}$.}
\item{There is a timescale that characterises the population of CRRs, such that they are largest (smallest in number) at $\tau_\mathrm{crr}\sim\tau_\alpha^{0.8}$.} 
\item{The fractal dimension of CRRs increases upon supercooling below $T_\mathrm{mct}$.}
\item{The CRR size distribution exhibits a power--law scaling, excluding rare, percolating CRRs.}
\item{At low temperature, particles in excitations are very frequently in CRRs. This is not the case at high temperature $T\gtrsim T_\mathrm{mct}$ where there is rather little correlation between particles in excitations and those in CRRs.}
\item{At low temperature, there is a huge decoupling in timescales of excitations which remains microscopic and in fact \emph{decreases} somewhat and that of CRRs which is coupled to the relaxation time. If we estimate the number of particles which have been in excitations during the CRR timescale, we find that this massively exceeds the number of particles in the system, suggesting that particles may participate many times in excitations if they are in a CRR. Conversely, many CRR particles may not satisfy the criteria for an excitation. The massive decoupling of timescales between CRRs and excitations makes it computationally challenging to probe whether a CRR particle is also in one (or more) excitations.}
\end{itemize}

\section*{Supplementary Material}
\setcounter{figure}{0}
\renewcommand\thefigure{\thesection S\arabic{figure}}

\textit{Simulation and model details. --- }
We study the binary Lennard-Jones Kob-Andersen (KA) mixture at the non-standard 2:1 and 3:1 composition in the NVT ensemble (Nose-Hoover thermostat) at $\rho = 1.400$ using the Roskilde University Molecular Dynamics (RUMD) package \cite{bailey2017}. We also simulate KA 4:1 for comparing our algorithms with data in the literature. The interatomic interactions of the KA mixture are defined by $v_{ij}(r) = \epsilon_{\alpha   \beta}\big[(\frac{\sigma_{\alpha \beta}}{r})^{12} - (\frac{\sigma_{\alpha \beta}}{r})^{6}]$ with parameters $\sigma_{AB} = 0.80$, $\sigma_{BB} = 0.88$ and $\epsilon_{AB} = 1.50$, $\epsilon_{BB} = 0.50$ ($\alpha$, $\beta$ = A, B). The pair potential is cut-and-shifted at $r_{c}$ = $2.5\sigma_{\alpha \beta}$. We employ a unit system in which $\sigma_{AA}$ = 1, $\epsilon_{AA}$ = 1, and $m_{A} = m_{B}$ = 1 (we refer to this as MD or LJ units). We study system sizes $N$ = 10,002 and 100,002 for KA 2:1 and $N$ = 10,000 and 100,000 for KA 3:1.  
The time step is $\Delta t$ = 0.0025 or 0.0035 depending on the temperature. The temperature is in the range $T$  = [0.48, 2.00] for KA 2:1 and $T$ = [0.68, 2.00] for KA 3:1, and at the lowest temperature required well over 1 year to equilibrate on a Nvidia 1080 GTX card. We confirmed equilibrium by running the simulations back-to-back at least twice and comparing the intermediate scattering functions.

\begin{table}[]
	\caption{Parameter $\alpha$ from the RFOT fit Eq.\,\ref{eqRFOTFit} and $T_{MCT}$ estimated with a power law fit Eq.\,\ref{eqMCTFit}.}
	\begin{tabular}{ccc}
		\hline
		& KA 2:1 & KA 3:1 \\ \hline
	$\alpha$	& 0.61 & 0.62 \\
	$T_{MCT}$	& 0.56  & 0.75 \\ \hline
	\end{tabular}
\end{table}

\begin{figure}
\includegraphics[width=1\linewidth]{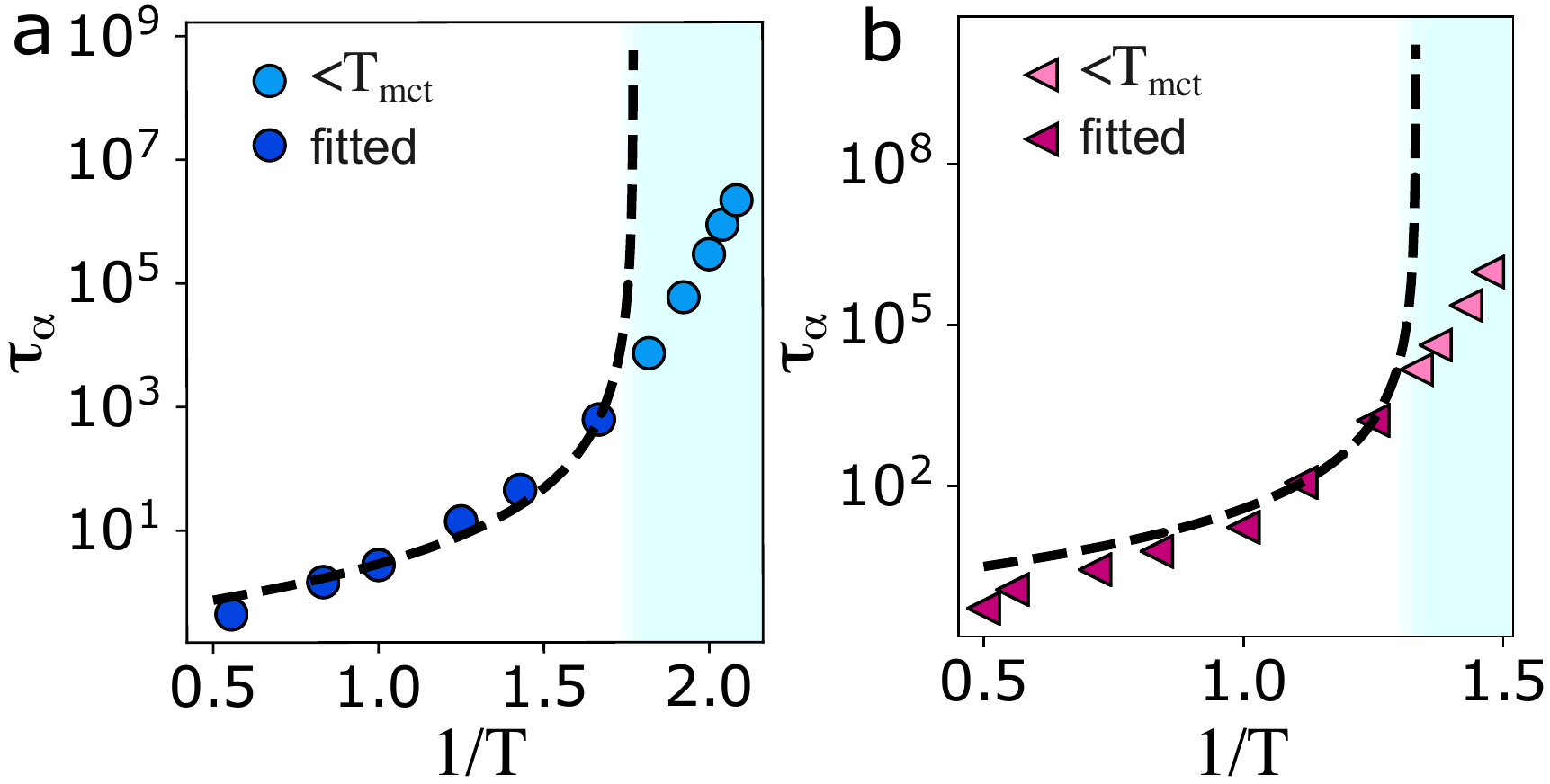}
\caption{
Power law fit of the relaxation time to estimate $T_{\rm mct}$. The plot is fitted with Eq.\ref{eqMCTFit}. Since the data is expected to go beyond $T_{\rm mct}$ some data had to be excluded from the fit as indicated.}
\label{sFigMCT}
\end{figure}

The mode-coupling temperature is around $T_{\rm mct}$ = 0.55 for KA 2:1 (see Fig. \ref{sFigMCT}).
The mode coupling temperature $T_\mathrm{mct}$ is estimated with a power law fit \cite{kob1995}.
	
\begin{equation}
\label{eqMCTFit}
\tau_{\alpha}= B\left(\frac{1}{T_\mathrm{mct}}-\frac{1}{T}\right)^{-\gamma}
\end{equation}

\begin{table}[]
	\caption{Parameters fitted to the various systems considered.}
	\begin{tabular}{ccccc}
		\cline{3-5}
		&  & \multicolumn{1}{|c|}{KA 2:1} & \multicolumn{1}{c|}{KA 3:1} & \multicolumn{1}{c|}{Experiments} \\ \hline
		\multicolumn{1}{|c|}{\multirow{3}{*}{VFT fit}} &   \multicolumn{1}{c|}{ $T^*$ or $Z^*$ }                    &         0.33           &  0.48                  &          \multicolumn{1}{c|}{  36.5}               \\
		\multicolumn{1}{|c|}{}                  &       \multicolumn{1}{c|}{$\tau_0$}                   &    0.05                   &   0.12                    &   \multicolumn{1}{c|}{ 0.39 }  \\
		\multicolumn{1}{|c|}{}                  &    \multicolumn{1}{c|}{$D$  }                  &       7.38                &        6.69               &      \multicolumn{1}{c|}{0.15}               \\ \cline{1-5}
		\multicolumn{1}{|c|}{\multirow{3}{*}{parabolic fit}} &   \multicolumn{1}{c|}{ $T_0$ or $Z_0$  }                   &   0.744                    &     1.16                  &               \multicolumn{1}{c|}{11.8}         \\
		\multicolumn{1}{|c|}{}                  &    \multicolumn{1}{c|}{$\tau_0$ }                   &    55.1                   &     24.6                  &   \multicolumn{1}{c|}{1.35}   \\
		\multicolumn{1}{|c|}{}                  &    \multicolumn{1}{c|}{$J$  }                   &                 4.46      &     5.35                  &         \multicolumn{1}{c|}{0.24}              \\ \cline{1-5}
	\end{tabular}
	\label{sTableParameters}
\end{table}

\textit{Experimental. --- }
Fluorescent dyed-poly methylmethacrylate (PMMA) colloids with a polyhydroxystearic acid comb stabiliser were synthesised using established methods \cite{elsesser2011}. To enhance spatial resolution between particles, a non-fluorescent PMMA shell was grown on the fluorescent cores and cross-linked with ethylene glycol dimethacrylate (EGDM) to yield a total radius of 270 nm (polydispersity $\approx$ 8\%, Brownian time $ \tau_B=34$ ms). PMMA particles were imaged in a density-- and refractive index matching mixture of cis-decalin and cyclohexyl bromide.

Samples were loaded into cells constructed of three coverslips on a microscope slide, where two of the coverslips acted as a spacer, and sealed using epoxy. The structural relaxation time was monitored for waiting times of up to 100 days until it reached a steady state at which point the sample was considered to be equilibrated \cite{elmasri2009}. For example, the state point $\phi=0.598$ reached equilibrium after 16 days, corresponding to more than $10 \tau_{\alpha}$. Samples were imaged using a Leica SP8 inverted stimulated emission depletion microscope with a 100x oil immersion lens in STED-3D mode, mounted on an optical table. Measurements were taken at least 20 particle diameters from the cell wall to minimise any structural or dynamic influence of surface layering, which typically persisted for around 5 particle diameters. Images were deconvolved with Huygens Professional version 15.05 prior to analysis (Scientific Volume Imaging, The Netherlands, \url{http://svi.nl}). 
Particle tracking algorithms \cite{leocmach2013} were used to determine particle coordinates from deconvolved STED microscopy data. 

\begin{figure}
\includegraphics[width=1.0\linewidth]{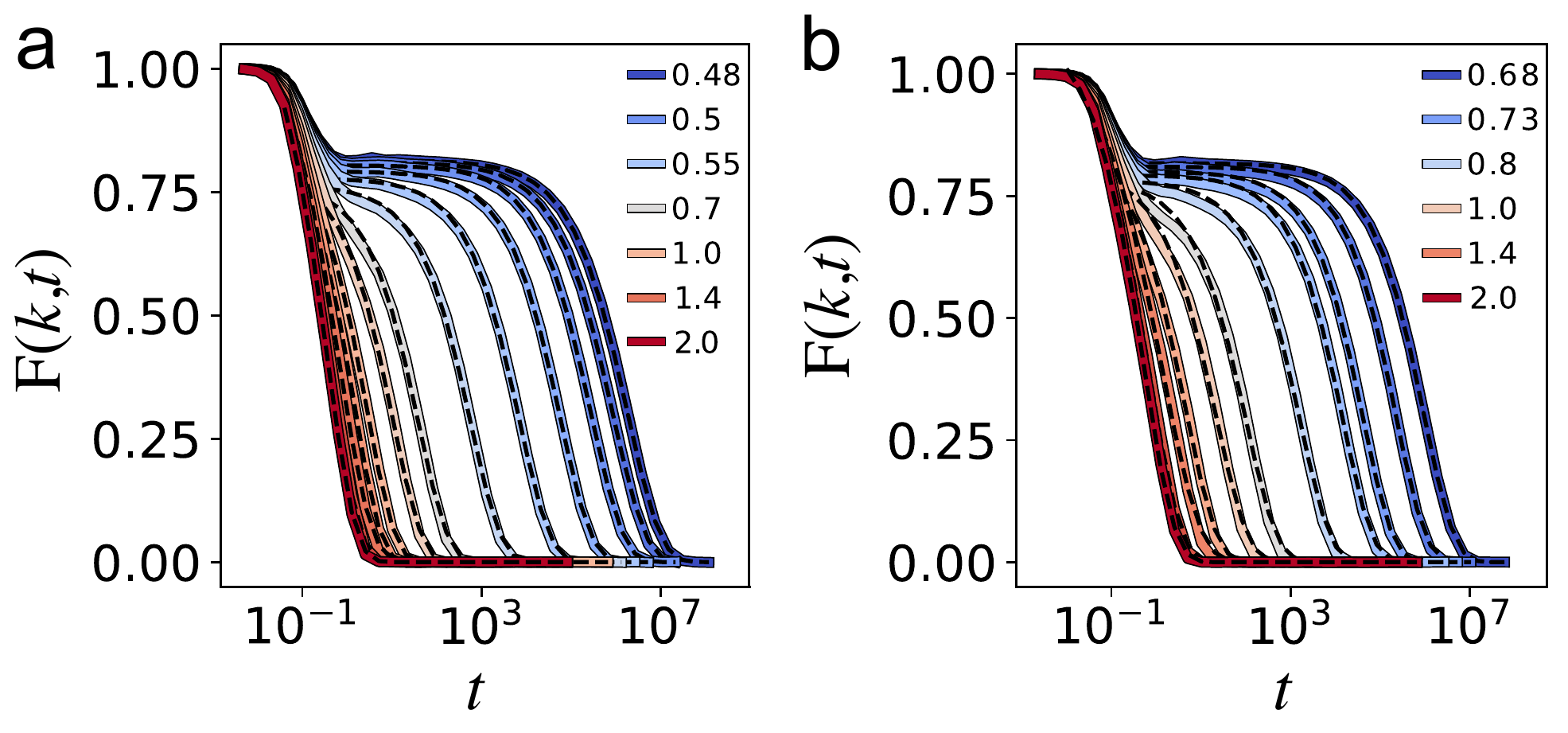}
\caption{ISFs fitted a stretched exponential for (a) KA 2:1 and (b) KA 3:1. Coloured lines are data, dashed black lines are fits.}
\label{SISF_fit}
\end{figure}

\begin{figure}
\includegraphics[width=1.0\linewidth]{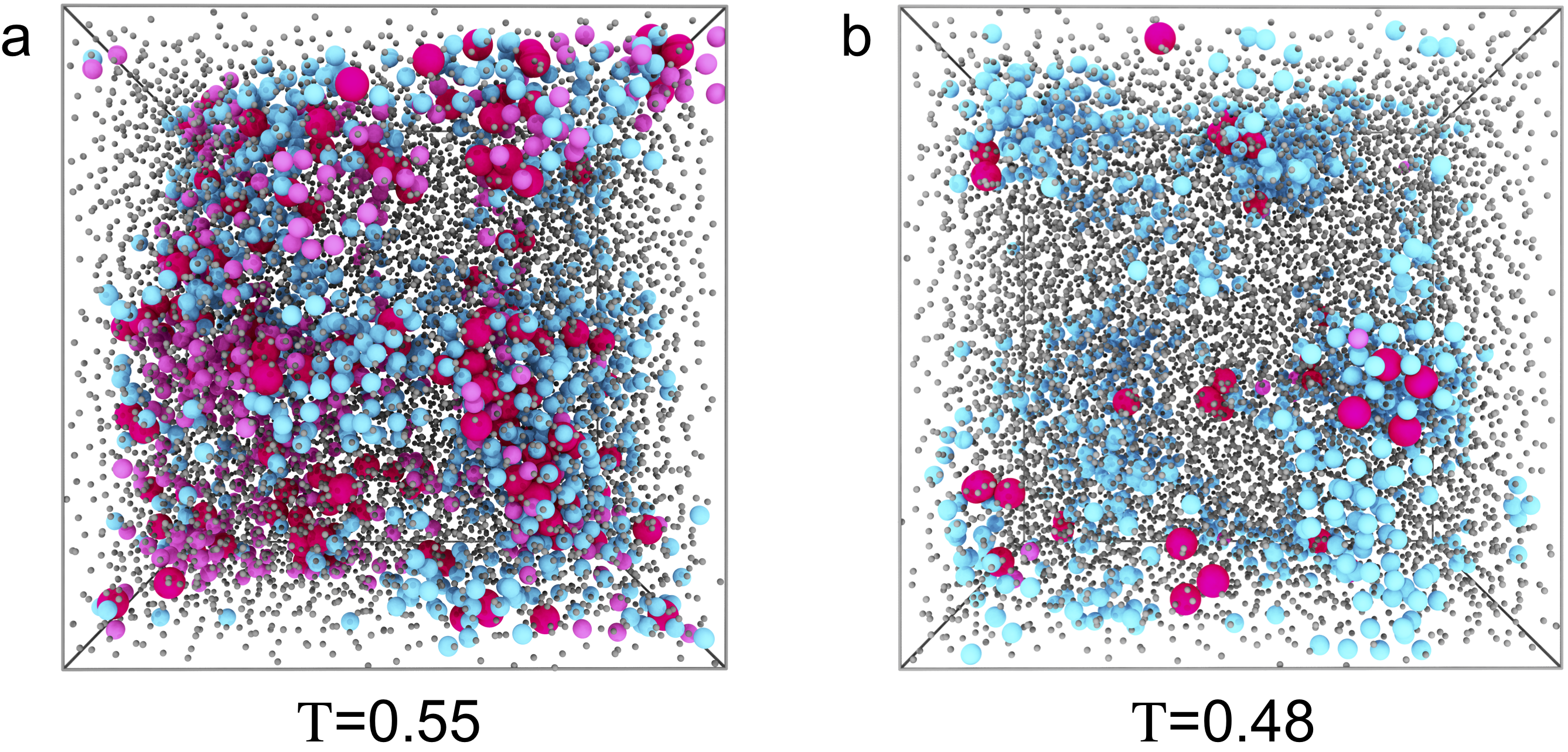}  
\caption{Snapshots of the KA 2:1 system and $T=0.55$ (a) and $T=0.48$ (b). 
Bright pink: particles both CRR and excitations. Blue: CRR. Pale pink: excitations. Grey: neither. Particles are not rendered to scale.
(b) is reproduced from Fig. \ref{figAngell}(a).                         
\label{sFigSnapshotT055}}
\end{figure}

\begin{figure}
\includegraphics[width=1.0\linewidth]{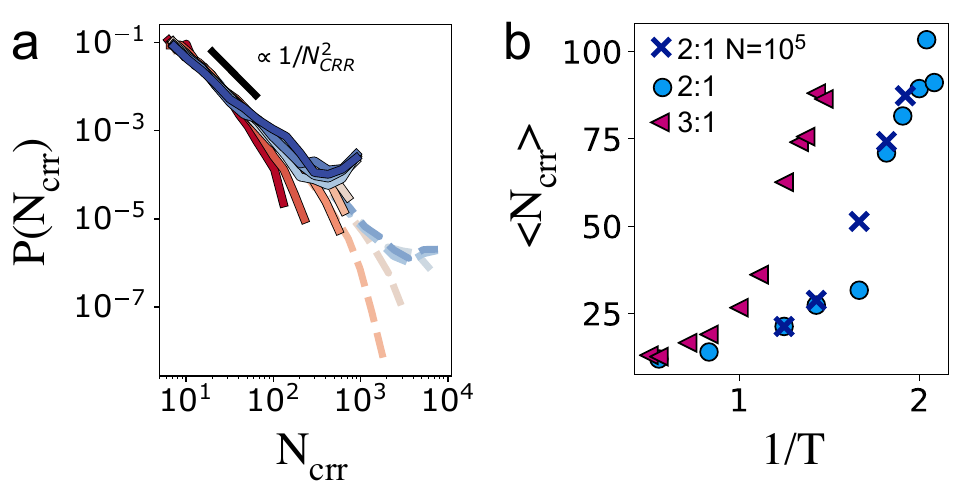}  
\caption{
(a) CRR size distribution of KA 2:1 model for $ N=10 002$ particles. Dashed lines are simulations with $N =100 002$ particles. 
(b) average number of particles in a cluster.} 
\label{sFigDistributionNCRR}
\end{figure}




\begin{figure}
	\includegraphics[width=0.5\linewidth]{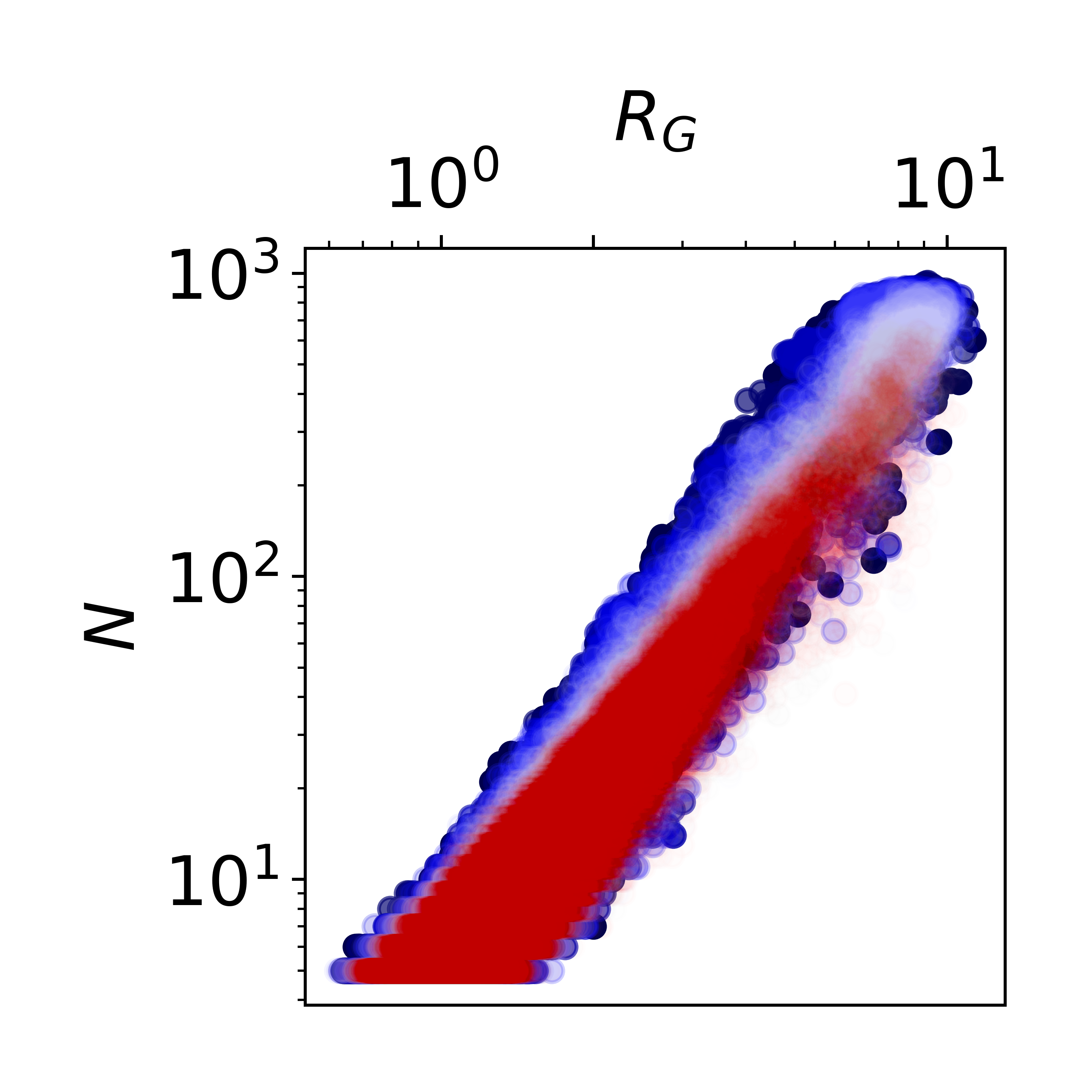}
	\caption{Number of particles in clusters as a function of their radius of gyration. The fractal dimension is the exponent of power law fit.}
	\label{sFigFractal}
\end{figure}



\begin{figure}
\includegraphics[width=0.5\linewidth]{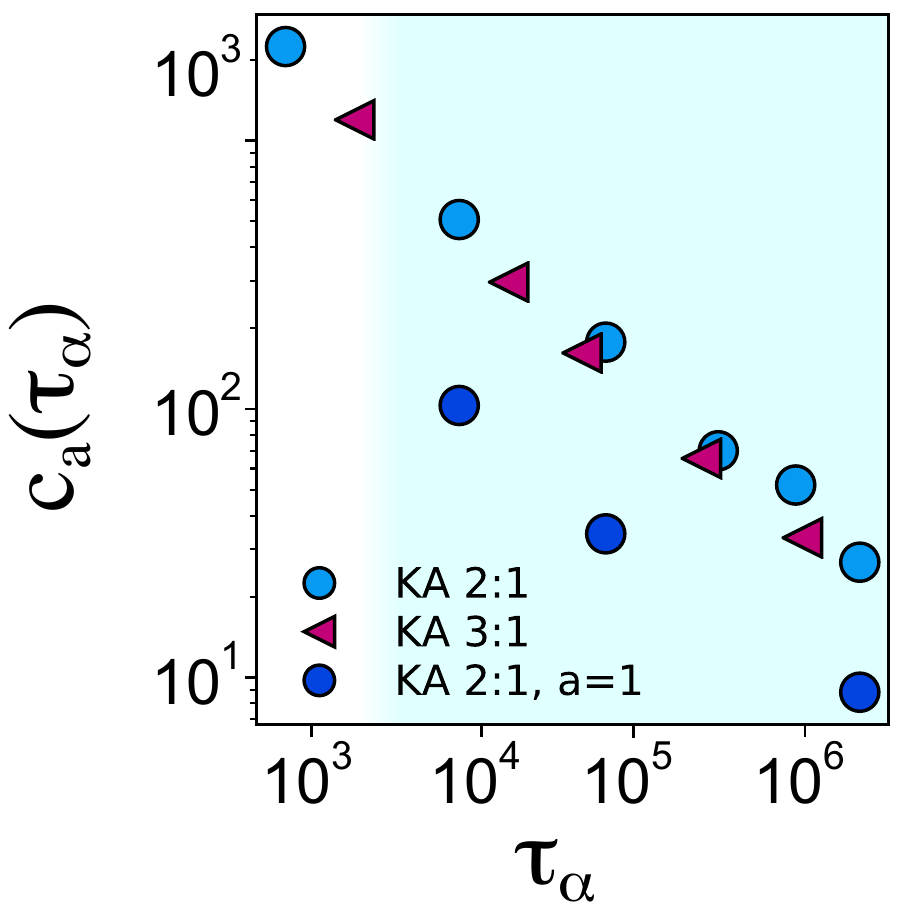}
\caption{Average number of excitations in a trajectory of length $t_\mathrm{traj}=1000$LJ units. }
\label{sFigExRaw}
\end{figure}

\begin{figure}
\includegraphics[width=0.5\linewidth]{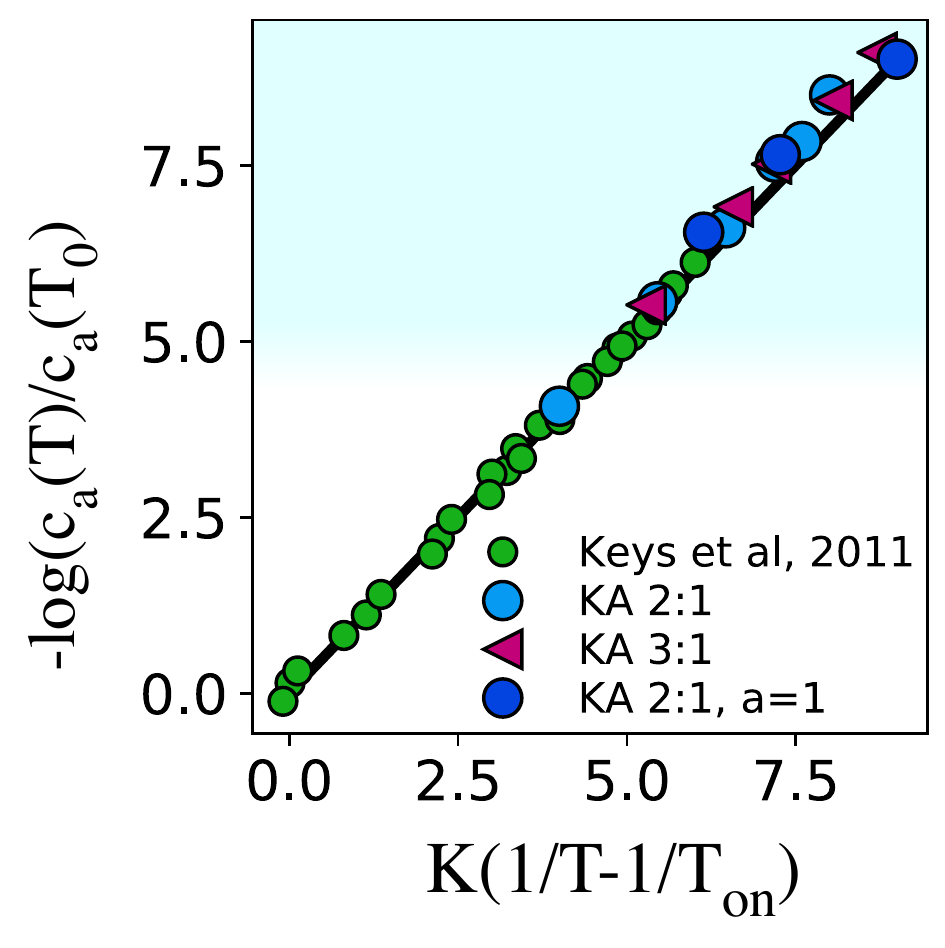}
\caption{Scaled temperature dependence of the concentration of excitations in comparison with the data by Keys et al. \cite{keys2011}. Same figure as Fig. 2(b) in the main manuscript but including excitation detection with $a=1$ for some temperatures}
\label{sFigExa1}
\end{figure}

\vspace{20pt}
\textit{Structural Relaxation Time and Dynamic susceptibility}

The relaxation time $\tau_{\alpha}$ is calculated
from the self-part of intermediate scattering function $F_S(k,t)$ for a wave vector near the first peak of the $A$-particle static structure factor. $F_S(k,t)$ is fitted with a stretched exponential:
\begin{equation}
F_{S}(k,t)=A\exp\left[\left(-\frac{t}{\tau_{\alpha}}\right)^{\beta}\right]
\end{equation}

\begin{align}
\chi_{4}(t) \equiv \frac{1}{N}\big(\langle q(t)^{2}\rangle - \langle q(t)\rangle^{2} \big)
\end{align}
The dynamic susceptibility is calculated from the self-part of the overlap function where
\begin{align}
q(t) = \sum_{i=1}^{N} \Theta\big(r_{a} - |\textbf{r}_{i}(t+t_{0}) - \textbf{r}_{i}(t_{0})|\big)
\end{align}
Here $\Theta$ is the Heaviside step function and $r_{a}$ = 0.30. The same value is used for both particles.

\vspace{20pt}
\textit{Configurational entropy --- } 
The configurational entropy $S_{\rm conf}$ is calculated from the total entropy $S$ by subtracting the basin entropy
in the harmonic approximation \cite{berthier2019,shila}

\begin{align}
S_{\rm conf} = S - S_{\rm vib}
\end{align}
where

\begin{align}
S_{\rm vib} = \Big\langle \sum_{a = 1}^{3N}(1-\ln(\beta \hbar \omega_{\rm a}) \Big\rangle_{\rm IS}
\end{align}
In this equation $\beta = 1/k_{\rm B}T$ and $\omega_{\rm a}^{2}$ is an eigenvalue of the Hessian matrix for the inherent state. The average is taken over 200 inherent states with about $\tau_{\alpha}$ in between the configurations. Inherent states are found by a simple gradient descent method which works well for the KA liquid.

The total entropy $S$ is calculated via thermodynamic integration from  low density (the ideal gas limit). Details of the calculation can be found in Ref. \cite{bell2020}. The definition of the total entropy in MD units requires a value for Planck's constant in MD units. We used $h^{*}$ = 0.1857 which is calculated from using $\sigma$ = 0.3405$\cdot 10^{-9}$ m and $\epsilon$ = 0.03994 kg/mol when converting to MD units \cite{shila}. 

The diffusion coefficient is determined from the long-time limit of the $A$-particle mean-square displacement.\\

\begin{figure}
\includegraphics[width=0.5\linewidth]{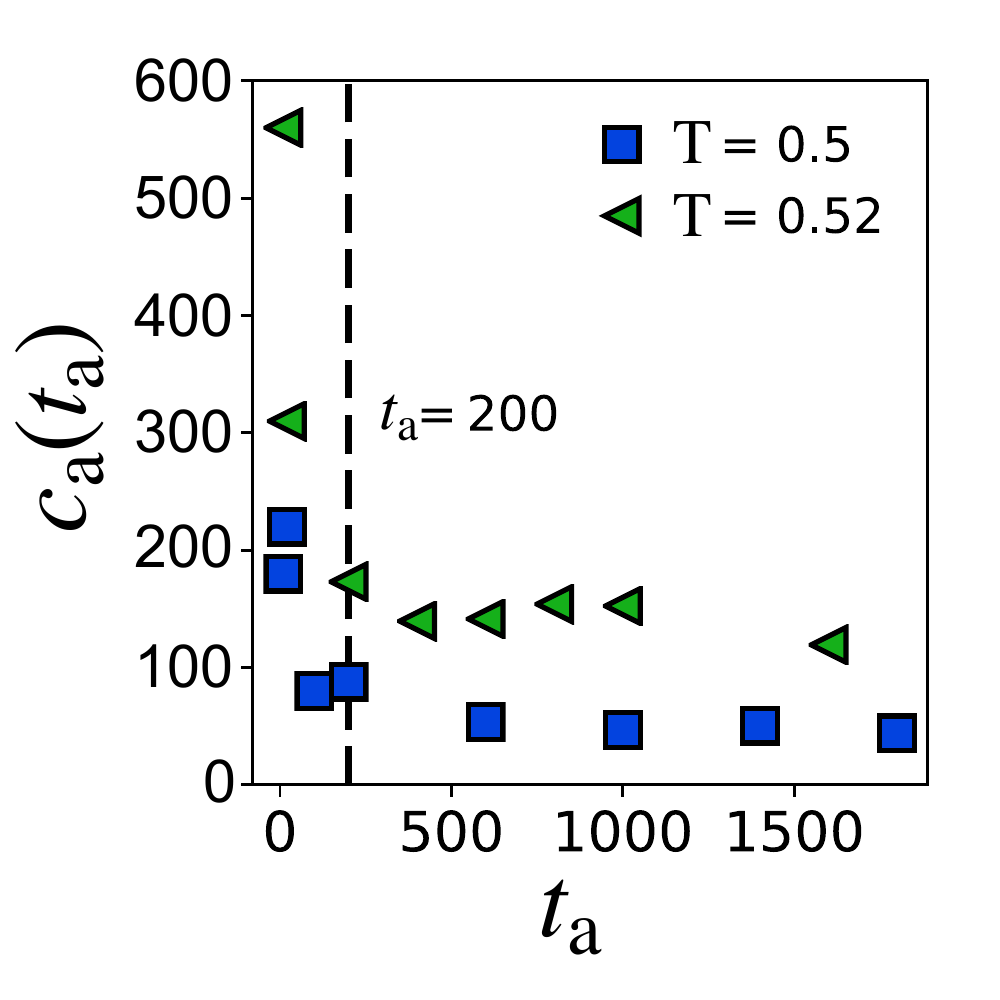}
\caption{Detected excitations depending on the choice of $t_a$ the lowest value $t_a=200$ in the constant region is chosen.}
\label{SFindTa}
\end{figure}

\vspace{10pt}
\textit{Excitation Detection Algorithm --- }
To identify the excitations the timescale $t_a$ and the length scale $a$ have to be chosen. $a$ has to be small enough that particles can jump the distance but large enough to exclude small fluctuation. After setting $a=0.5$ the algorithm is used for different $t_a$. Fig. \ref{SFindTa} shows that there is a range where the concentration of excitations does not depend on $t_a$. The smallest $t_a$ in this range was chosen ($t_a=200\,$LJ units).

Trajectories of length $5t_a$ containing 1000 frames are generated and each particle is tested as a potential excitation. For this the average particle positions of the first and the  last $t_a$ of the trajectory are compared and if this is smaller than $a$ the particle is rejected. This tests if the particle has committed to a new position (step 1).

If the particle has not been rejected we check for each frame if the average positions in the previous $t_a$ and the following $t_a$ are at least $a$ apart. This makes sure that we exclude particles that move slowly and steadily since we do not consider these to be excitations. This is important at higher temperatures but almost irrelevant at lower temperatures (step 2).
To get a first estimate for the hyperbolic tangent fit, for every identified frame $f$ that fulfills the above condition, we go forwards and backwards in time and identify the closest frame for which the distance of the position for the current frame is larger than $a/2$  (step 3). The time between those two frames is the first guess for $\Delta t$ and the central frame of the excitation that produces the shortest $\delta t$ guess serves as an input for the centre of the fit (step 4). The trajectory is smoothed with a spline (step 5 )and a fitted with a hyperbolic tangent function to extract the exact centre of the excitation and $\delta t$ (step 6). If this fit fails, the trajectory is rejected as well. This algorithm starts failing around $T_\mathrm{mct}$ since fall particles move a lot and excitations aren't clearly defined. For low temperatures well below  $T_\mathrm{mct}$  this algorithm is computationally not too expensive since most particles are already rejected in the first step so that the more costly procedure (moving window, fit) are only preformed on very few particles.

\begin{enumerate}
	\item 
	for all particles if: $\frac{1}{t_a}\left(\sum\limits_{t=4t_a}^{5t_a}\mathbf{x}(t)-\sum\limits_{t=0}^{t_a}\mathbf{x}(t)\right)<a$
	reject
	\item else: for every frame f: 
	if $\frac{1}{t_a}\left(\sum\limits_{t=f-t_a}^{f}\mathbf{x}(t)-\sum\limits_{t=f}^{f+t_a}\mathbf{x}(t)\right)>a$
	\item for every frame $f$ fullfilling these criteria: \newline
	$\delta_{\text{temp}} =\min(\mathbf{x}(t))_{\mathbf{x}(t)-\mathbf{x}(f)>a/2;t}-\max(\mathbf{x}(t))_{\mathbf{x}(f)-\mathbf{x}(t)>a/2;t<f}$
	\item use frame with $\min{\delta_{\text{temp}}}$ as initial guess for $\tanh$ fit
	\item smooth trajectory with spline
	\item fit trajectory with $\tanh$ to get $\delta t$ or reject if fit fails
\end{enumerate}

\begin{acknowledgments}
We thank Ludovic Berthier, Giulio Biroli, Rob Jack, Camille Scalliet, and Shiladitya Sengupta
for insightful suggestions and discussions. CPR and FT acknowledges the European Research Council (ERC consolidator grant NANOPRS, project 617266) and the Engineering and Physical Sciences Research Council (EP/H022333/1) for financial support. TSI is supported by the VILLUM Foundation’s Matter grant (No. 16515).
\end{acknowledgments}


%

\end{document}